\documentclass[conference]{IEEEtran}
\IEEEoverridecommandlockouts
\usepackage{amsmath,amssymb,amsfonts}
\usepackage{graphicx}
\usepackage{textcomp}
\usepackage{xcolor}

{}
{}
{}

\usepackage{adjustbox}
\usepackage{multirow}
\usepackage{color, colortbl}

\definecolor{LightCyan}{rgb}{0.88,1,1}
\definecolor{lightgreen}{rgb}{0.80,1.0,0.80}
\definecolor{lightgray}{rgb}{0.83, 0.83, 0.83}

\newcommand{\MM}{\mathcal{M}}

\newcommand{\indep}{\perp \!\!\! \perp}

\newcommand{\ie}{\textit{i}.\textit{e}. }

 \renewcommand{\baselinestretch}{0.92}

\usepackage{algorithm}
\usepackage[noend]{algpseudocode}

\usepackage[font=footnotesize]{caption} 
\usepackage{subfig}
\usepackage{booktabs} 
\usepackage{multirow}
\usepackage{diagbox}
\usepackage[compress]{cite}

\renewcommand{\baselinestretch}{0.93}
\def\BibTeX{{\rm B\kern-.05em{\sc i\kern-.025em b}\kern-.08em
    T\kern-.1667em\lower.7ex\hbox{E}\kern-.125emX}}

\begin{document}
\bstctlcite{IEEEexample:BSTcontrol}


\title{Diverse Knowledge Distillation (DKD): A Solution for Improving The Robustness of Ensemble Models Against Adversarial Attacks
}

\author{\IEEEauthorblockN{Ali Mirzaeian$^*$ , Jana Kosecka$^*$, Houman Homayoun$^\dagger$, Tinoosh Mohsenin$^\ddagger$, Avesta Sasan$^*$ }
$^*$Department of ECE, George Mason University, e-mail: \{amirzaei,  kosecka, asasan\}@gmu.edu\\
$^\dagger$Department of ECE, University of California, Davis, e-mail: hhomayoun@ucdavis.edu\\
$^\ddagger$Department of ECE, University of Maryland, Baltimore County, e-mail: tinoosh@umbc.edu
}

\maketitle

\begin{abstract}
This paper proposes an ensemble learning model that is resistant to adversarial attacks. To build resilience, we introduced a training process where each member learns a radically distinct latent space. Member models are added one at a time to the ensemble. Simultaneously, the loss function is regulated by a reverse knowledge distillation, forcing the new member to learn different features and map to a latent space safely distanced from those of existing members.  We assessed the security and performance of the proposed solution on image classification tasks using CIFAR10 and MNIST datasets and showed security and performance improvement compared to the state of the art defense methods.
\end{abstract}

\vspace{2mm}

\section{Introduction}\label{sec:introduction}
In the past decade, the research on Neuromorphic-inspired computing models, and the applications of Deep Neural Networks (DNN) for estimation of hard-to-compute functions, or learning of hard-to-program tasks have significantly grown, and their accuracy have considerably improved. Early research on learning models mostly focus on improving the accuracy of the models \cite{alexnet, resnet}, but as models matured, researchers explored other dimensions, such as energy efficiency of the models \cite{icnn, exploit, neshatpour2019icnn, hosseini2019complexity, 8119201, neshatpour2018design} and underlying hardware \cite{nesta, eyeris, mirzaeian2019tcd,  10.1145/3427377, daneshtalab2020hardware, Faraji_ISCAS_2020, 9116473}, as well as wider adoption and application of learning solutions in many other research fields including security applications \cite{vakil2020lasca, sayadi20192smart, vakil2020learning} for solving problems that either have no closed-form solution or are too complex for developing a programmable solution. The wide adoption of these capable solutions then started raising concerns over their security.  

Among many security aspects of learning solutions, the vulnerability of models to adversarial attacks  \cite{szegedy2013intriguing, Goodfellow2014}, has attracted lots of attention.  Researchers have shown that subtle, yet targeted adversarial perturbation to the input (\ie image, audio, or video input) of neural networks can dramatically drop their performance \cite{ papernot2016limitations,Papernot2016PracticalBA}.

The vulnerability of DNNs to adversarial attacks has raised serious concerns for using these models in critical applications in which an adversary can slightly perturb the input to fool the model \cite{Elsayed2018AdversarialET}. This paper focuses on adversarial attacks on image classification models where an adversary manipulates an input image, forcing the DNN to misclassify.

Non-robust features are those features that strongly associate within a specific class, yet have small variation across categories \cite{CodeBridged}. Ilyas and et al. at \cite{Ilyas2019AdversarialEA} showed that the high sensitivity of the underlying model to the non-robust features existing at the input dataset is a significant reason for the vulnerability of the model to the adversarial examples. So an adversary crafts a perturbation that accentuates the non-robust features to achieve a successful adversarial attack.

From this discussion, a means of building robust classifiers is identifying robust features and training a model using only robust features (that have a sizeable intra-class variation), making it harder for an adversary to mislead the classifier, \cite{ Ilyas2019AdversarialEA}. Motivated by this discussion, we proposed a simple yet effective method for improving the resilience of DNNs, by introducing auxiliary model(s) trained in the spirit of knowledge distillation, while forcing diversity across features formed in their latent spaces.

\begin{figure}[t!]
    \begin{center}
    \includegraphics[width=0.70\columnwidth]{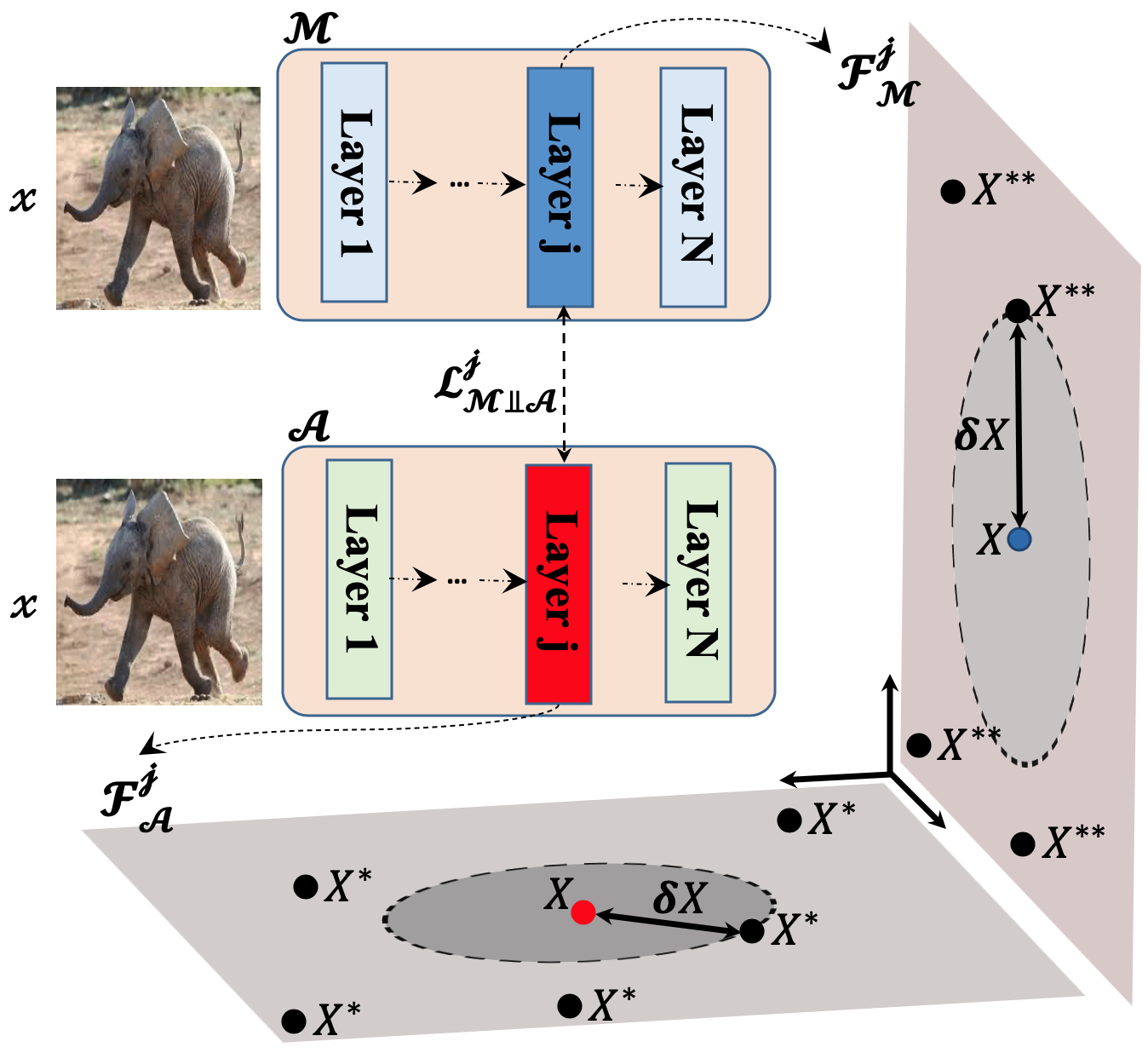}
    \end{center}
    \caption{In our proposed solutions, we train an auxiliary model(s) that tracks the main model's classification while learning a diverse set of features, latent representation of which maps to space far apart from the teacher's. $\mathcal{L}_{\MM \indep \mathcal{A}}^{j}$ indicates the diversity between the $j^{th}$ feature space of the model $\mathcal{M}$ and $\mathcal{A}$  which are $\mathcal{F}_{\mathcal{M}}^{j}$ and $\mathcal{F}_{\mathcal{A}}^{j}$, respectively. }
    \label{fig:outlie}
\end{figure}

We argue that adversarial attacks are transferable across models because they learn similar latent spaces for non-robust features. This is either the result of using the 1) same training set or 2) knowledge distillation that solely focused on improving the classification accuracy. In other words, for sharing the dataset or the knowledge of a trained network (on a dataset), the potential vulnerabilities of models coincide. Hence, an attack that works on one model is very likely to work on the other(s). This conclusion is also supported by the observations by Ilyas and et al. at \cite{Ilyas2019AdversarialEA}. From this argument, we propose to augment the task of knowledge distillation with an additional and explicit requirement that the learned features by the student (\ie auxiliary) model(s) should be distinct and independent from the teacher (\ie main) model. For this reason, as showed in Fig. \ref{fig:outlie}, we introduce the concept of Latent Space Separation (LSS), forcing the auxiliary model to learn features with little or no correlation to those of the teacher's.  Hence, an adversarial attack on the main model will have minimal impact on the latent representation of features learned by the auxiliary model(s).

\section{Background}
Prior researches on adversarial learning have produced different explanations on why leaning models are easily fooled by adversarial input perturbation. The early investigations blamed the non-linearity of neural networks for their vulnerability \cite{Goodfellow2014,biggio2013evasion}. However, this perception was later challenged by Goodfellow and et. al. \cite{Goodfellow2014}, who developed the Fast Gradient Sign Method (FGSM), explaining how neural network linearity can be exploited for rapidly building adversarial examples. 

Building robust learning models that could resist adversarial examples has been a topic of interest for many researchers. Some of the most notable prior art on this topic includes 1) Adversarial Training, 2) Knowledge Distillation (KD), and 3) de-noising and refinement of the adversarial examples. 

\begin{figure}[h!]
    \centering
    \includegraphics[width=0.75\columnwidth]{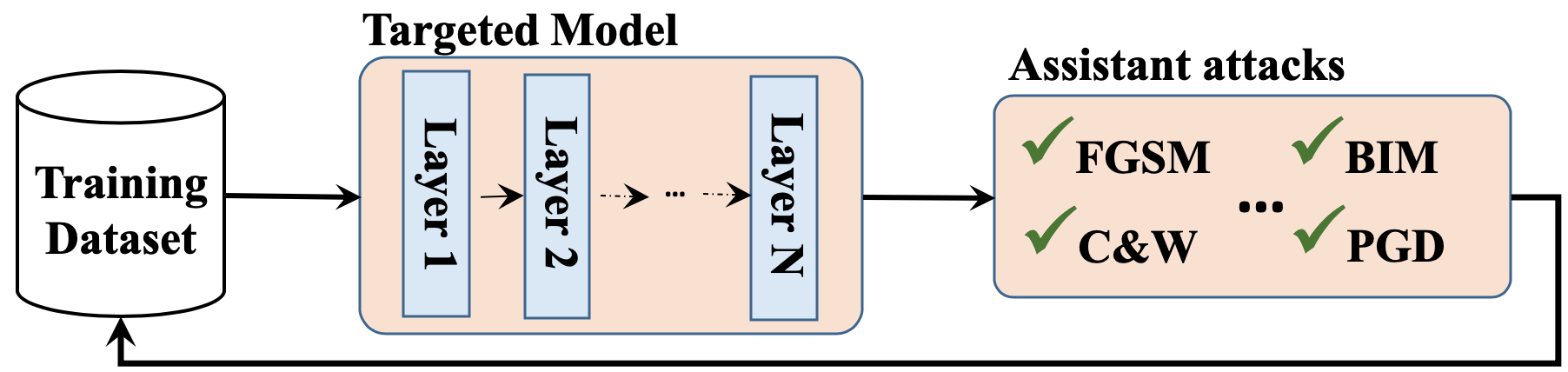}
    \caption{Adversarial training in a nutshell.}
    \label{fig:adv_training}
\end{figure}

\textbf{Adversarial Training}: It is the process of incremental training of a model with the known adversarial examples to improve its resilience, see Fig\ref{fig:adv_training}. The problem with this approach is that the model's resilience only improves when the model is attacked with similarly generated adversarial examples \cite{sai, Tramr2019AdversarialTA}. 

\begin{figure}[h!]
    \centering
    \includegraphics[width=0.75\columnwidth]{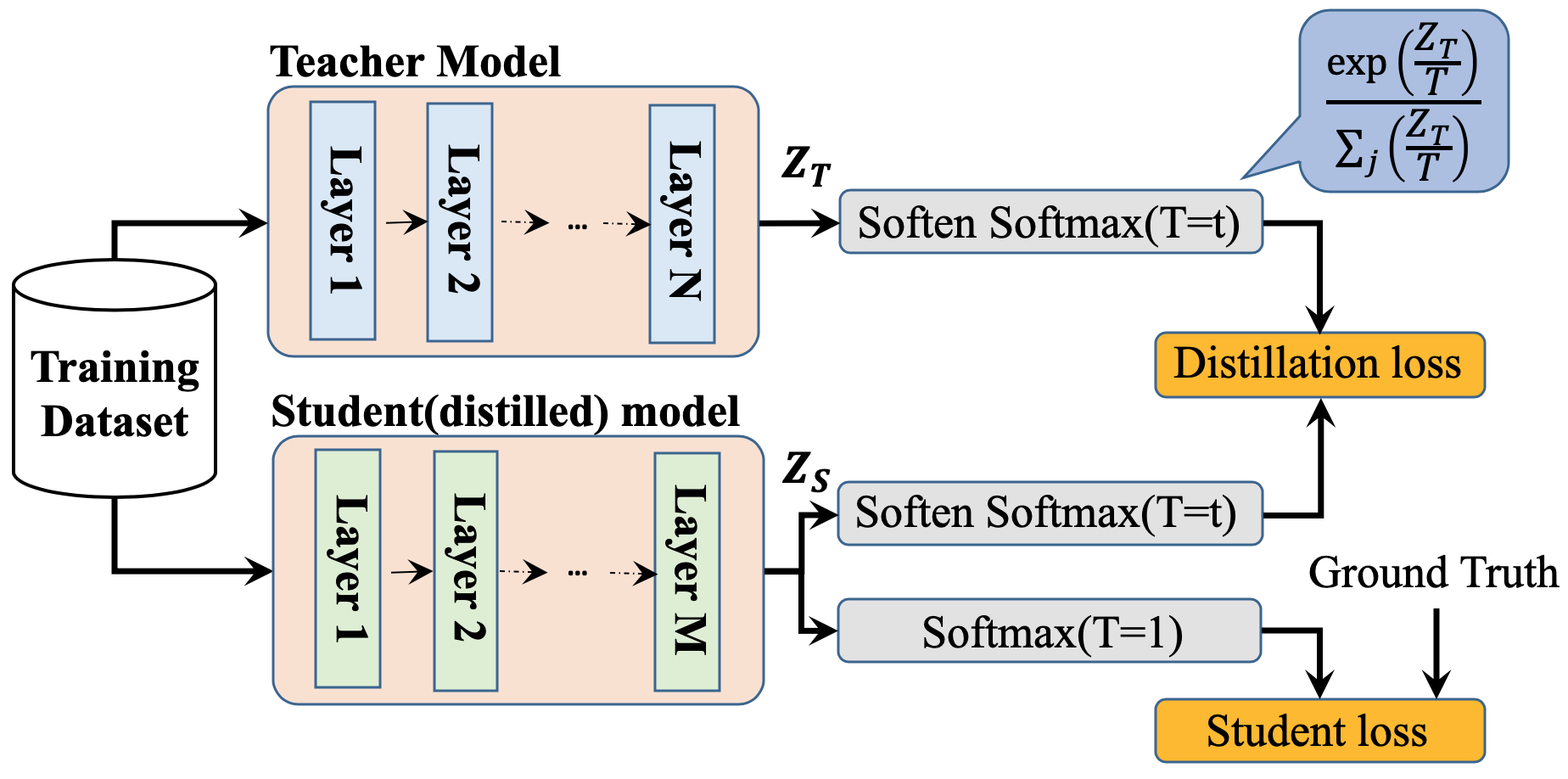}
    \caption{Defensive Distillation in a nutshell.}
    \label{fig:def_dist}
 \end{figure}

\textbf{Knowledge Distillation (KD)}, see Fig. \ref{fig:def_dist}: In this method, a compact (student) model learns to follow one or more teacher's models behavior. It was originally introduced to build compact models (students) from more accurate yet larger models (teachers). Later, it was also used to diminish the sensitivity of the student's output model concerning the input's perturbations \cite{Papernot2015DistillationAA}. However, the work in \cite{carlini2016defensive} showed that if the attacker has access to the student model, with minor changes, the student model could be as vulnerable as the teacher. Specifically,  knowledge distillation can be categorized as a gradient masking defense \cite{pmlrv80athalye18a} in which the magnitude of the underlying model's gradients are reduced to minimize the effect of changes in the model's input to its output. Although grading masking defenses can be an effective defense against white-box attacks, they are not resistant against black-box evasion attacks  \cite{carlini2016defensive}. Our proposed solution is motivated by KD. However, we do not force the auxiliary (\ie student) network(s) to follow the output layers of the main network (\ie teacher); in contrary to the KD, the auxiliary network has to learn a different latent space while being trained for the same task and on the same dataset. 

\begin{figure}[h!]
    \centering
    \includegraphics[width=0.75\columnwidth]{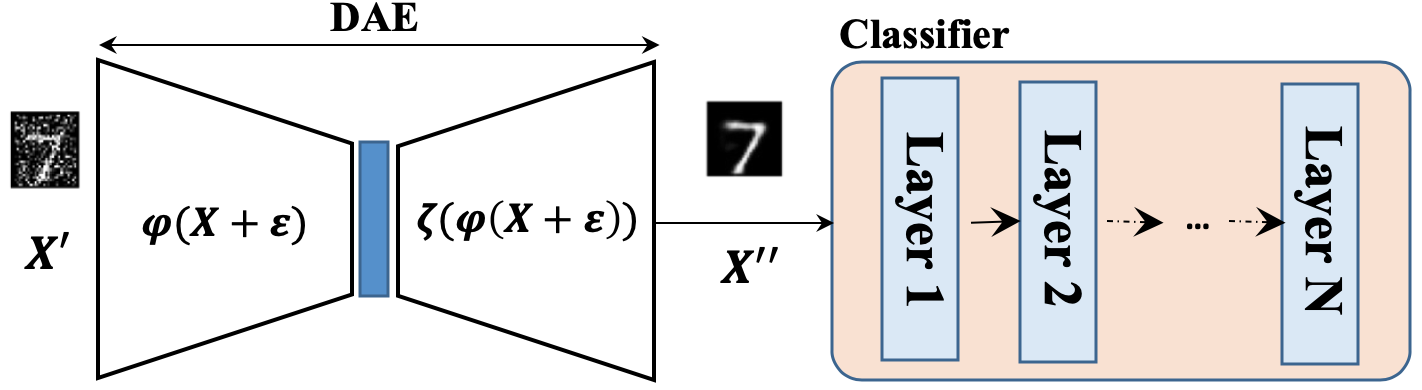}
    \caption{Autoencoders for defense.}
    \label{fig:dae}
\end{figure}

\textbf{Refining the Input Image}, see Fig \ref{fig:dae}: The adversarial defenses that rely on refining the input samples try to denoise the input image using some sort of autoencoder (variational, denoising, etc.)  \cite{comparative}. In this approach, the image is first encoded using a deep network to extract a latent code (a lossy, compressed, yet reconstructable representation of the input image). Then the image is reconstructed using a decoder. Next, the decoded image is fed to the classifier  \cite{comparative}. However, this approach suffers from two main weaknesses: (1) The reconstruction error of decoder can significantly reduce the classifier's accuracy and such reconstruction error increases as the number of input classes increases; (2) the used encoder-network is itself vulnerable to adversarial attacks which means new adversarial examples can be crafted on the model which also include the encoder-network.

\section{The Proposed Method}

Our objective is to formulate a knowledge distillation process in which one or more auxiliary (student) models $\mathcal{A}_i$ are trained to closely follow the prediction of a main (teacher) model $\MM$, while being forced to learn substantially different latent spaces. For example, in Fig. \ref{projection}, let's assume three auxiliary models $\mathcal{A}_1$, $\mathcal{A}_2$, $\mathcal{A}_3$ have been trained alongside the main model $\mathcal{M}$ to have the maximum diversity between their latent space representations. Our desired outcome is to assure that an adversarial perturbation that could move the latent space of the input sample $x$ out of its corresponding class boundary, $\mathcal{F}_{\mathcal{M}}^{j}$, has a negligible or small impact on the movement of the corresponding latent space of $x$ in the class boundaries of the auxiliary models $\mathcal{A}_{1}$, $\mathcal{A}_{2}$, $\mathcal{A}_{3}$. Hence, an adversarial input that could fool model $\MM$, becomes less effective or ineffective on the auxiliary models. This objective is reached by the way the loss function of each model is defined. The details of $\MM$ and $\mathcal{A}_i$ network(s), learning procedure and objective function of $\mathcal{A}_i$ are explained next:

\begin{figure}[hbt!]
    \centering
    \includegraphics[width=0.75\columnwidth]{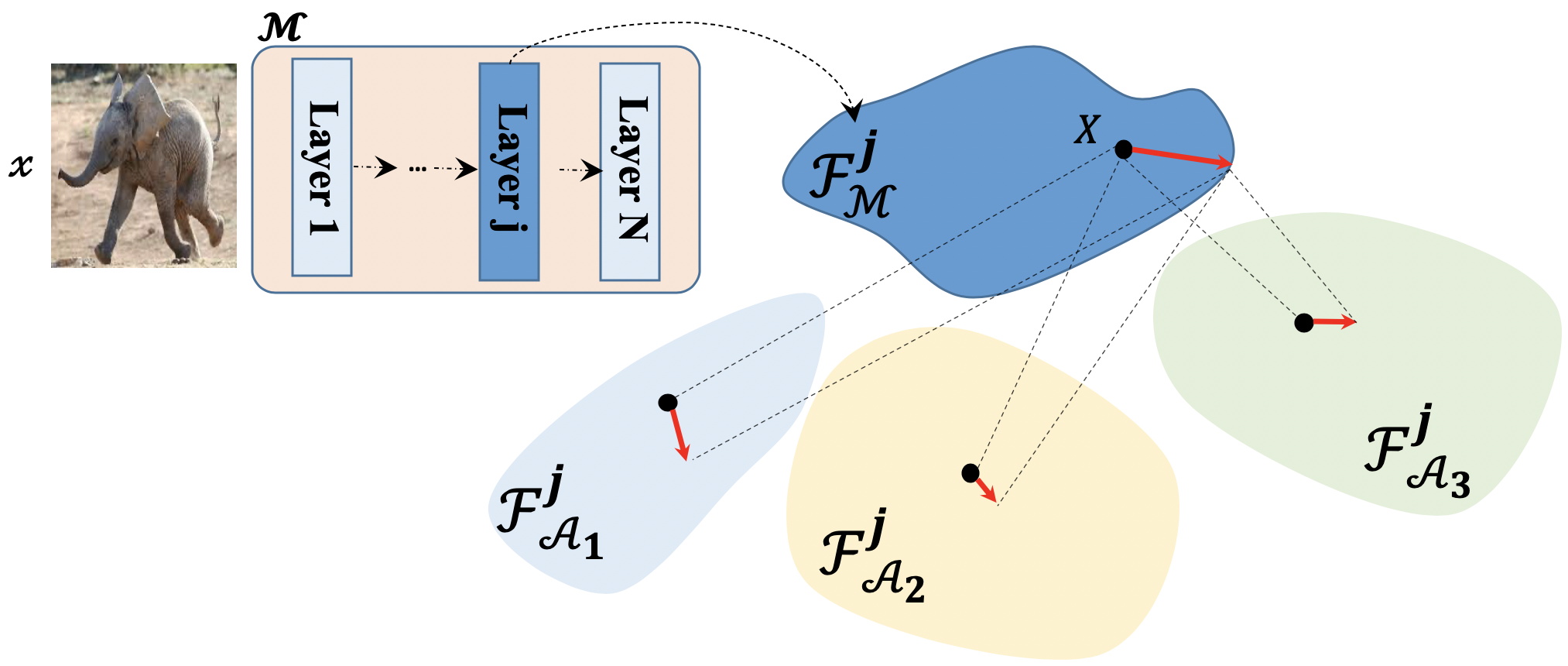}
    \caption{$\mathcal{F}_{\mathcal{M}}^{j}$ is the latent space regards to j$^{th}$ layer of the main model, $\mathcal{F}_{\mathcal{A}_{1}}^{j}$, $\mathcal{F}_{\mathcal{A}_{2}}^{j}$, $\mathcal{F}_{\mathcal{A}_{3}}^{j}$ are the corresponding latent spaces of the auxiliary models 1, 2, and 3. $X$ is the latent feature of an input sample $X$ i.e., $X = \mathcal{F}_{\mathcal{M}}^{j}(x) $. The red arrow at the main model shows the direction of the adversarial perturbation, the red arrows for the auxiliary models show the projection of the perturbation on the latent spaces of the auxiliary model.}
    \label{projection}
\end{figure} 
 
\textbf{Main Model ($\MM$:)}  In this paper, we evaluated our proposed method on two datasets MNIST \cite{lecun1998gradient} and CIFAR10 \cite{krizhevsky2009learning}. So depending on the underlying dataset, the structure of the main model  (teacher)  $\MM$ is selected as showed in Table \ref{architecture}. We also employed cross-entropy loss $\mathcal{C}$, see Eq. \ref{eq:entropy}, as the objective function for training the model $\MM$. 

\begin{equation}
    \small
    \mathcal{L}_\MM=\mathcal{C}(Y,\MM(X))=-\sum_{i=1}^{i=N}{Y_i \log(\MM(X_i))}
    \label{eq:entropy}
\end{equation}
\normalfont

In this equation, $X_i$ and $Y_i$ are $i^{th}$ training sample and its corresponding label, respectively. 

\textbf{Auxiliary Model ($\mathcal{A}$):} Each auxiliary model is a structural replica of the main model $\MM$. However, model $\mathcal{A}$ is trained using a modified KD training process: let's denote the output of $j^{th}$ layer of model $\mathcal{A}$ (\ie latent space of model $\mathcal{A}$) and $\MM$ by  $\mathcal{F}_\mathcal{A}^{j}$ and $\mathcal{F}_\mathcal{M}^{j}$ respectively. Our training objective is to force model $\mathcal{A}$ to learn very different latent space compared to $\MM$ where both do the same classification task on the same dataset. To achieve this, the term $\mathcal{L}_{\MM \indep \mathcal{A}}^{j}$ which shows the similarity of the $j^{th}$ layer of model A  to $j^{th}$ layer of model $\MM$ is defined as follows: 

\small
\begin{equation}
   \mathcal{L}_{\MM \indep \mathcal{A}}^{j}=
   \frac{\mathcal{F}(X)_\mathcal{M}^{j}\,\,\textbf{.}\,\, \mathcal{F}(X)_\mathcal{A}^{j}}{|\mathcal{F}(X)_\mathcal{M}^{j}||\mathcal{F}(X)_\mathcal{A}^{j}|} ,\quad X \in \mathbf{T}
\end{equation}
\normalfont

In this objective function, $\mathbf{T}$ is the dataset, and the $(.)$ is the inner product function. This similarity measure then is factored to define the loss function $(\mathcal{L}_\mathcal{A})$ for training  the model $\mathcal{A}$, which increases the dissimilarity of the layer $j$ of the model $\mathcal{A}$ with respect to the $\MM$:  

\small
\begin{equation}
    \mathcal{L}_\mathcal{A}=\underbrace{(1-\zeta)\mathcal{L}_\mathcal{C}(Y,\mathcal{A}(X))}_{targets\, the\, accuracy}+ \underbrace{\zeta \mathcal{L}_{\MM \indep \mathcal{A}}^{j}}_{targets\, the \, diversity}
    \label{eq:loss_a}
\end{equation}
\normalfont

In this equation, $\zeta$ is a regularization parameter to control the contribution of each term. 

Let's assume the adversarial perturbation $\delta$, when added to the input $X$, forces the model $\MM$ to misclassify $X$, or more precisely $\MM(X+\delta) \neq \MM(X)$. For this misclassification (evasion) to happen, in a layer $j$ (close to the output) the added noise has forced some of the class-identifying features outside its related class boundary learned by model $\MM$.  However, the class boundaries learned by $\mathcal{A}$ and $\MM$ are quite different. Therefore, as showed in Fig. \ref{projection}, although noise $\delta$ can move a feature out of its learned class boundary in model $\MM$, it has very limited power in displacing the features learned by model $\mathcal{A}$ outside of its class boundary in layer $j$. In other words, although the term $\mathcal{F}(X)_\mathcal{M}^{j} \,\, \textbf{.} \,\, \mathcal{F}(X)_\mathcal{A}^{j}$ between the main and the auxiliary models  has a low value, the term $\mathcal{F}(X+\delta)_\mathcal{A}^{j} \,\, \textbf{.} \,\, \mathcal{F}(X)_\mathcal{A}^{j}$ between the auxiliary model before and after adding perturbation $\delta$ has a high value, subsequently the student model $\mathcal{A}$ has a low sensitivity to the perturbation $\delta$, meaning $\mathcal{A}(X+\delta)=\mathcal{A}(X)$. 

\subsection{Black and White-Box defense:}
The auxiliary models could defend against both white and black box attacks, description and explanation for each is given next:

\textbf{Black-box Defense:}
In black box attack, an attacker has access to model $\MM$, and can apply her desired input to the model and monitor the model's prediction for designing an attack and adding the adversarial perturbation to the input $X$. Considering no access to the model $\mathcal{A}$, and for having very different feature space, the model $\mathcal{A}$ remains resistant to black-box attacks and using a single $\mathcal{A}$ is sufficient. 

\textbf{White-box Defense:} In white-box attack, the attacker knows everything about the model $\MM$ and $\mathcal{A}$, including model parameters and weights, full details of each model's architecture, and the dataset used for training the network. For this reason, using a single model $\mathcal{A}$ is not enough, as that model could be used for designing the attack. However, we can make the attack significantly more difficult (and improve the classification confidence) by training and using multiple robust auxiliary models. However, each of our $\mathcal{A}$ models learns different features compared with all other auxiliary models. Then, to resist the white-box attack, we create a majority voting system from the robust auxiliary models. 

Let's assume we want to train $k > 1$ auxiliary models,  $\mathcal{A}_{i=1}^{i=k}$, each having a diverse latent space (\ie $\MM \indep \mathcal{A}_1 \indep \mathcal{A}_2 \indep \cdots \indep \mathcal{A}_k $).  To learn these networks,  firstly, based on Eq. \ref{eq:loss_a}, the $\mathcal{A}_1$ is learned aiming $\MM \indep \mathcal{A}_1$. Then, the $\mathcal{A}_2$ is learned to be diverse of both $\mathcal{A}_1$ and $\MM$. This process continues one by one, reaching the $\mathcal{A}_i$ model, till its latent space is diverse from all previous models (\ie  $\mathcal{A}_i \indep \{\MM, \mathcal{A}_{j=1}^{j=i-1}\}$). According to this discussion for learning the i$^{th}$ auxiliary model, the loss function is defined as: 
\small     
\begin{equation}\label{eq:loss_b}
    \mathcal{L}_\mathcal{A} = (1-\zeta)\mathcal{L}_\mathcal{C}(Y,\mathcal{A}_i(X))+  (\zeta/i)\mathcal{L}_{\MM \indep \mathcal{A}_i}^{i} 
    +(\zeta/i)\sum_{j=1}^{y=i-1}\mathcal{L}_{ \mathcal{A}_y \indep \mathcal{A}_i}^{j}
\end{equation}    
\normalfont

Finally, to increase the confidence of the prediction, instead of a simple majority voting system for the top-1 candidate, we consider a boosted defense in which the voting system considers the top $n$ candidates of each model $\mathcal{A}$ for those cases that majority on top-1 fails (there is no majority between the top-1 predictions). This gives us two benefits 1) if a network misclassifies due to adversarial perturbation, there is still a high chance for the network to assign a high probability (but not the highest) to the correct class. 2) if a model is confused between a correct class and a closely related but incorrect class and assigns the top-1 confidence to the wrong class, it still helps identify the correct class in the voting system.

\section{Experimental Results}
In this section, we evaluate the performance of our proposed defense against various white and black-box attacks. We trained all models using Pytorch framework\cite{NEURIPS2019_9015}, and all attack scenarios using Foolbox\cite{rauber2017foolbox} (which is a toolbox for crafting adversarial examples). The details of the hyperparameters and system configuration are summarized in Table \ref{architecture}.


\begin{table}
    
    \centering
    \caption{The Architecture of the ensemble models}
    \begin{adjustbox}{width=0.8\columnwidth,center}
    	\begin{tabular}{|c|c|c|c|}
    		\hline 
    		\multicolumn{2}{|c|}{\textbf{MNIST Architecture} } & \multicolumn{2}{c|}{\textbf{CIFAR10 Architecture} }\tabularnewline
    		\hline 
    		\hline 
    		Relu Convolutional & 32 filters (3$\times$3) & Relu Convolutional  & 96 filters (3$\times$3)\tabularnewline 
    		Relu Convolutional & 32 filters (3$\times$3) & Relu Convolutional  & 96 filters (3$\times$3)\tabularnewline 
    		Max Pooling        & 2$\times$2              & Relu Convolutional  & 96 filters (3$\times$3)\tabularnewline 
    		Relu Convolutional & 64 filters (3$\times$3) & Max Pooling         & 2$\times$2\tabularnewline              
    		Relu Convolutional & 64 filters (3$\times$3) & Relu Convolutional  & 192 filters (3$\times$3)\tabularnewline 
    		Max Pooling        & 2$\times$2              & Relu Convolutional  & 192 filters (3$\times$3)\tabularnewline 
    		Relu Convolutional & 200 units                                  & Relu Convolutional  & 192 filters (3$\times$3)\tabularnewline 
    		Relu Convolutional & 200 units                                  & Max Pooling         & 2$\times$2\tabularnewline              
    		Softmax            & 10 units                                   & Relu Convolutional  & 192 filters (3$\times$3)\tabularnewline 
    		                   &                                            & Relu Convolutional  & 192 filters (1$\times$1)\tabularnewline 
    		                   &                                            & Relu Convolutional  & 192 filters (1$\times$1)\tabularnewline 
    		                   &                                            & Global Avg. Pooling & \tabularnewline                                           
    		                   &                                            & Softmax             & 10 units\tabularnewline    
    		\hline 
    		\hline 
    		\multicolumn{4}{|c|}{ \textbf{System Configuration and training hyper parameters} }\tabularnewline

    		\hline 
    		\multicolumn{4}{|p{1.09\linewidth}|}{OS: Red Hat 7.7, Pytorch: 1.3, FoolBox: 2.3.0,  GPU: Nvidia Tesla V100, EPOCH: 100,
    		MNIST Batch Size: 64, CIFAR10 Batch Size:128, Optimizer: ADAM, learning rate: 1e-4 }\tabularnewline
    		\hline 
    	\end{tabular}
	\end{adjustbox}
	\label{architecture}

\end{table}

\begin{figure}[ht!]
    \centering
    \includegraphics[width=0.85\columnwidth]{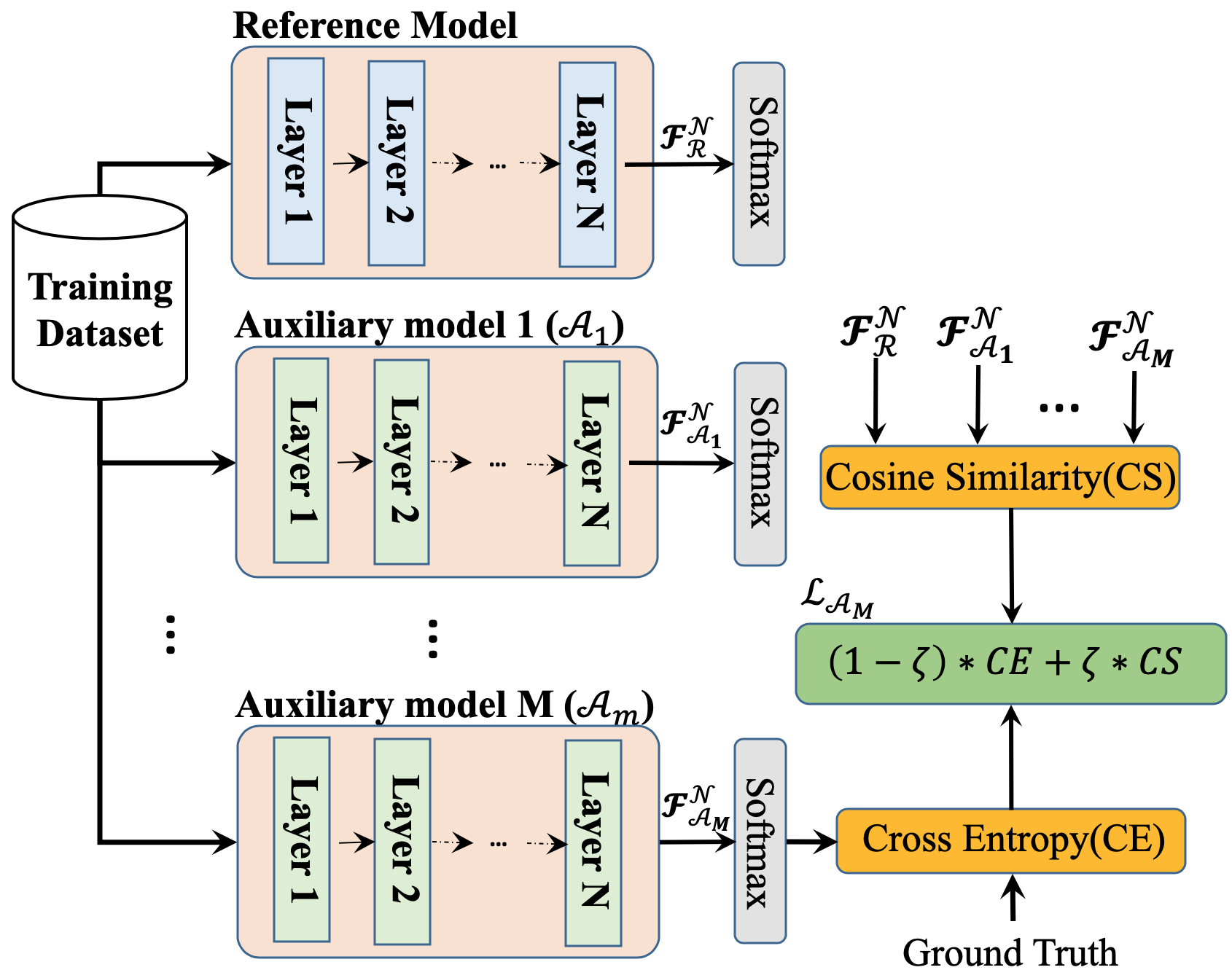}
    \caption{training method of the DKD for training an ensemble of m auxiliary models $\mathcal{A}_1$ to $\mathcal{A}_m$ from a reference model $\mathcal{R}$. $\mathcal{F}_{\mathcal{R}}^{\mathcal{N}}$, $\mathcal{F}_{\mathcal{A}_1}^{\mathcal{N}}$, and $\mathcal{F}_{\mathcal{A}_m}^{\mathcal{N}}$ are latent space of the reference model and auxilary models $\mathcal{A}_1$ and $\mathcal{A}_m$, related to output of $N^{th}$ layer of each one, respectively. Cross-Entropy \cite{Mannor2005TheCE} and Cosine-Similarity   are the objective functions that have been used for obtaining the total loss.}
    \label{blackbox}
\end{figure}

\subsection{Enforcing Latent Space Hetromorphisim}
To quantify the diversity of the latent space representations of the ensemble trained on a dataset $D$, we first define the Latent Space Separation (LSS) measure between latent spaces of two models $A_1$ and $A_2$ as:

\begin{equation}
    LSS_D(\mathcal{F}_{\mathcal{A}_{1}}^{j},                 \mathcal{F}_{\mathcal{A}_{2}}^{j}) = \frac{2}{\lVert W \rVert}.
    \label{sdm_equ}
\end{equation}

\begin{figure}[ht!]
    \centering
    \includegraphics[width=0.99\columnwidth]{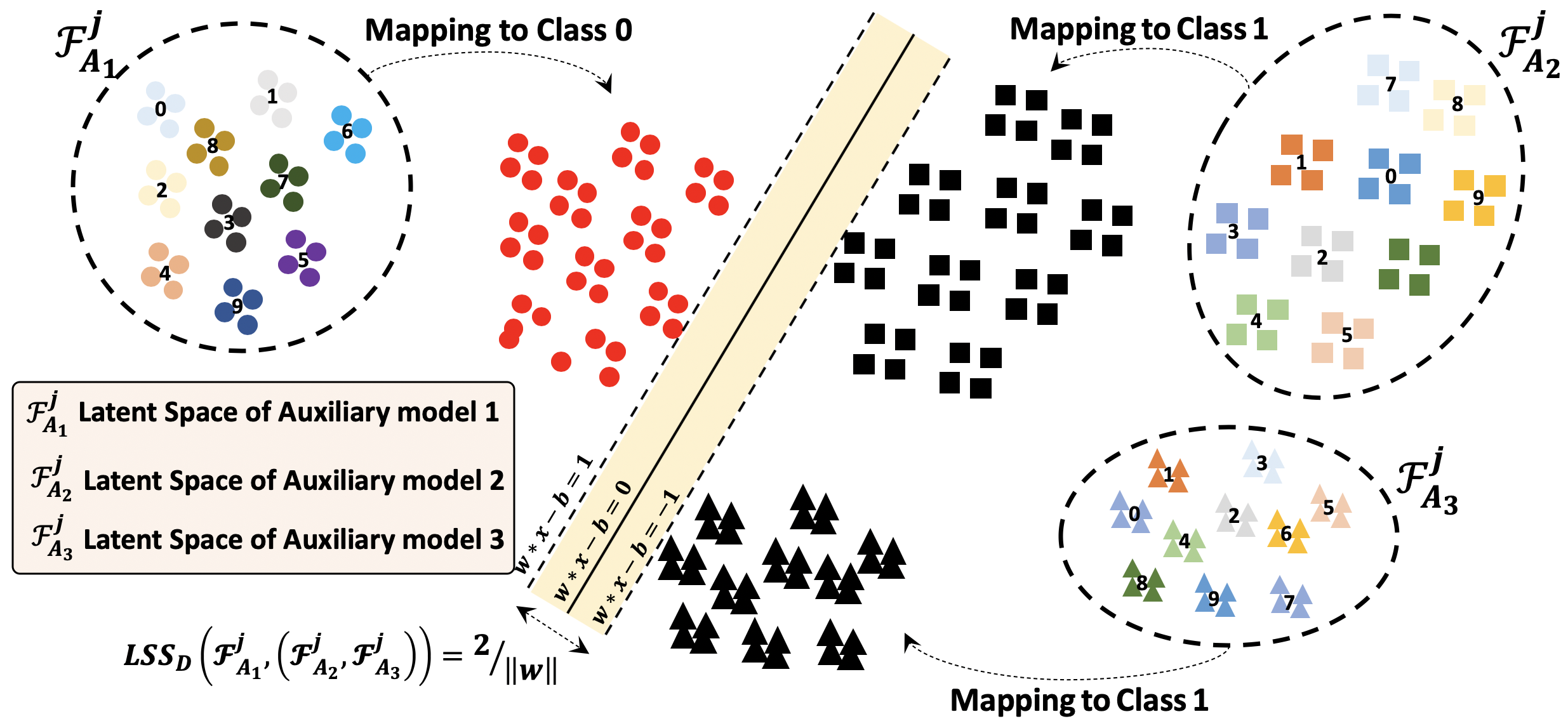}
    \caption{$LSS_D(\mathcal{F}_{\mathcal{A}_{1}}^{j},                 (\mathcal{F}_{\mathcal{A}_{2}}^{j}, \mathcal{F}_{\mathcal{A}_{3}}^{j} ))$ in which  $\mathcal{A}_{1}$, $\mathcal{A}_{2}$ and $\mathcal{A}_{3}$ are three models with latent space representations $\mathcal{F}_{\mathcal{A}_{1}}^{j}$, $\mathcal{F}_{\mathcal{A}_{2}}^{j}$, and $\mathcal{F}_{\mathcal{A}_{3}}^{j}$, respectively. $W$ is the norm vector of a hard margin support vector machine (SVM) in two class classification model. }
    \label{sdm2}
\end{figure}

In which $\mathcal{F}_{\mathcal{A}_{1}}^{j}$, $\mathcal{F}_{\mathcal{A}_{2}}^{j}$ are latent space representations of the dataset D for the j$^{th}$ layer of models $A_1$ and $A_2$, respectively. $W$ is the normal vector of the hyperplane obtained by Support Vector Machine (SVM) classifier\cite{cortes1995support} for linearly separate the latent spaces obtained on the dataset $D$. More precisely, LSS between two latent spaces is obtained by following these 4 steps: 1) training both models $\mathcal{A}_{1}$, $\mathcal{A}_{2}$ on a dataset i.e., MNIST.   2) generating the latent space of each model on the evaluation set i.e., $\mathcal{F}_{\mathcal{A}_{1}}^{j}$ and $\mathcal{F}_{\mathcal{A}_{2}}^{j}$. 3) turning the latent representations into a two-class classification problem tackled by SVM classifier  4) using SVM margin as LSS distance between two latent representations of the dataset MNIST, $LSS_{MNIST}(\mathcal{F}_{\mathcal{A}_{1}}^{j}, \mathcal{F}_{\mathcal{A}_{2}}^{j})$. 
This process can be expanded for obtaining LSS of more than two models. For instance Fig. \ref{sdm2} shows the LSS between model $\mathcal{A}_1$ for two models $\mathcal{A}_2$ and $\mathcal{A}_3$ regards to their j$^{th}$ layer, i.e., $LSS_D(\mathcal{F}_{\mathcal{A}_{1}}^{j},                 (\mathcal{F}_{\mathcal{A}_{2}}^{j}, \mathcal{F}_{\mathcal{A}_{3}}^{j} ))$. Note that the SVM classifier should be set in the hard margin mode, meaning no support vector can pass the margins. When the SVM classification fails, it means that the latent spaces were not linearly separable i.e., there is either an overlap between latent spaces or the decision boundary cannot be modeled linearly. So in both cases, the more the marginal distance between latent spaces are, the higher is the diversity of formed latent spaces.      

Using Eq.~\ref{sdm_equ}, we define more generalized formula of LSS using an ensemble model comprised of $N$ models, see Eq.~\ref{sdmn_equ}. In fact the total LSS of an ensemble model is obtained by averaging the LSS of each model latent space versus all other models'. For instance, let's imagine the ensemble model comprised of three models $A_1$, $A_2$, and $A_3$. Then the total LSS is calculated by $1/3 (LSS_D(F_{A_1}^j,(F_{A_2}^j, F_{A_3}^j))+LSS_D(F_{A_2}^j,(F_{A_1}^j, F_{A_3}^j))+ LSS_D(F_{A_3}^j,(F_{A_1}^j, F_{A_2}^j)) )$. LSS measures the marginal distance of SVM in a two-class-classification task, so $LSS_D(A_1,(A_2, A_3))$, indicates that a SVM classification has been performed between the latent space of the model $A_1$  and an aggregation of latent spaces of other two models $A_2$ and $A_3$.

\begin{equation}
    \frac{1}{N}\sum_{i=1}^{N}LSS_D(F_{A_i}^j, (F_{A_1}^j,..., F_{A_{i-1}}^j, F_{A_{i+1}}^j,..., F_{A_N}^j )).
    \label{sdmn_equ}
\end{equation}

\begin{table}[!htbp]
    \caption{The number of failed majority between an ensemble of the three models at the original and boosted version, indicated with *, of KD, DKD, and RI on the datasets MNIST and CIFAR10. The accuracy improvement using the boosted version is shown with Accuracy Improved (A.I). We investigated our proposed method against well-know attacks like DeepFool \cite{universal}, C\&W \cite{cw}, JSMA \cite{papernot2016limitations}, and FGSM \cite{Goodfellow2014}. }
    
	\centering
	\label{majority_fail}
	\begin{adjustbox}{width=0.98\columnwidth,center}
	\begin{tabular}{|c|c|c|c|c|c|c|c|c|c|c|c|c|}
		\hline 
		\multirow{2}{*}{\textbf{Param.}} & \multicolumn{6}{c|}{\textbf{MNIST}} & \multicolumn{6}{c|}{\textbf{CIFAR10}}\tabularnewline
		\cline{2-13} 
		     & \textbf{KD} & \textbf{KD{*}} & \textbf{DKD} & \textbf{DKD{*}} & \textbf{RI} & \textbf{RI{*}} & \textbf{KD} & \textbf{KD{*}} & \textbf{DKD} & \textbf{DKD{*}} & \textbf{RI} & \textbf{RI{*}}\tabularnewline 
		\hline
		\hline
		\rowcolor{lightgray}
		\multicolumn{7}{|c|}{Deep Fool} & \multicolumn{6}{c|}{Deep Fool}\tabularnewline
		\hline 
		1    & 4	&0	&15	&0	&16	&0    &  576	&6	&1234	&51	&891	&33 \tabularnewline 
		\hline 
		200  &  7	&1	&93	&0	&61	&0       &  628	&5	&1319	&52	&959	&32\tabularnewline 
		\hline 
		A.I.(\%) & \multicolumn{2}{c|}{0.02} & \multicolumn{2}{c|}{0.12} & \multicolumn{2}{c|}{0.35} & \multicolumn{2}{c|}{1.36} & \multicolumn{2}{c|}{2.88} & \multicolumn{2}{c|}{7.16}\tabularnewline
		\hline 
		\rowcolor{lightgray}
		\multicolumn{7}{|c|}{C\&W} & \multicolumn{6}{c|}{C\&W}\tabularnewline
		\hline 
		1    &   188	&3	&172	&0	&225	&6     & 584	&1	&1201	&2	&859	&13\tabularnewline 
		\hline 
		200  &     560	&5	&794	&2	&921	&16       &   630	&2	&1322	&2	&942	&12\tabularnewline 
		\hline 
		A.I.(\%) & \multicolumn{2}{c|}{0.01} & \multicolumn{2}{c|}{0.33} & \multicolumn{2}{c|}{0.22} & \multicolumn{2}{c|}{13.97} & \multicolumn{2}{c|}{1.47} & \multicolumn{2}{c|}{2.79}\tabularnewline
		\hline 
		\rowcolor{lightgray}
		\multicolumn{7}{|c|}{JSMA} & \multicolumn{6}{c|}{JSMA}\tabularnewline
		\hline 
		1    &    0	&0	&15	&0	&16	&6    & 584	& 8	& 1201	& 51	& 859	& 31\tabularnewline 
		\hline 
		200  &     4	&1	&52	&0	&42	&8   & 654	& 8	& 1410	& 56	& 999	& 35 \tabularnewline 
		\hline 
		A.I.(\%) & \multicolumn{2}{c|}{0.01} & \multicolumn{2}{c|}{0.26} & \multicolumn{2}{c|}{0.15} & \multicolumn{2}{c|}{1.39} & \multicolumn{2}{c|}{2.94} & \multicolumn{2}{c|}{4.18}\tabularnewline
		\hline 
		\rowcolor{lightgray}
		\multicolumn{7}{|c|}{FGSM} & \multicolumn{6}{c|}{FGSM}\tabularnewline
		\hline 
		0.04 &     190	&5	&342	&16	&235	&20   &  821	&3	&1801	&106	&1560	&66\tabularnewline 
		\hline 
		0.08 &    239	&6	&589	&27	&349	&34   &   924	&9	&1842	&71	&1872	&77 \tabularnewline 
		\hline 
		0.1  &   349	&3	&844	&36	&504	&46  &   950	&12	&1829	&79	&2053	&105\tabularnewline 
		\hline 
		A.I.(\%) & \multicolumn{2}{c|}{0.03} & \multicolumn{2}{c|}{0.21} & \multicolumn{2}{c|}{0.12} & \multicolumn{2}{c|}{1.33} & \multicolumn{2}{c|}{3.74} & \multicolumn{2}{c|}{2.94}\tabularnewline
		\hline 
	\end{tabular}
    \end{adjustbox}
\end{table}

To investigate the effectiveness of LSS, as a metric for measuring the diversity between different latent spaces, we considered three different scenarios for training an ensemble of 3 models with the same structure: I) Random Initialization (RI) where 3 models trained independently with a random initial value. II) Knowledge Distillation (KD), where three models trained collaboratively, as shown in Fig. III) Diversity Knowledge Distillation (DKD), where three models are trained in a collaborative yet different manner of KD, see Fig. \ref{blackbox}. Note that KD and RI methods deal with the softmax probabilities, However, DKD uses a mix of softmax and the latent spaces. Alongside each one of the designs DKD, KD, and RI, a boosted version of each one is implemented and denoted by $DKD^*$, $KD^*$, and $RI^*$, respectively. Both KD and DKD are trained in a one-by-one manner, meaning the $i^{th}$ model considers the $i-1$ previously trained models at its training phase while those $i-1$ models are frozen i.e., their parameters (weights) are not updated while the $i^{th}$ model is being trained.

\begin{figure*}[ht!]
    \centering
    \includegraphics[width=2\columnwidth]{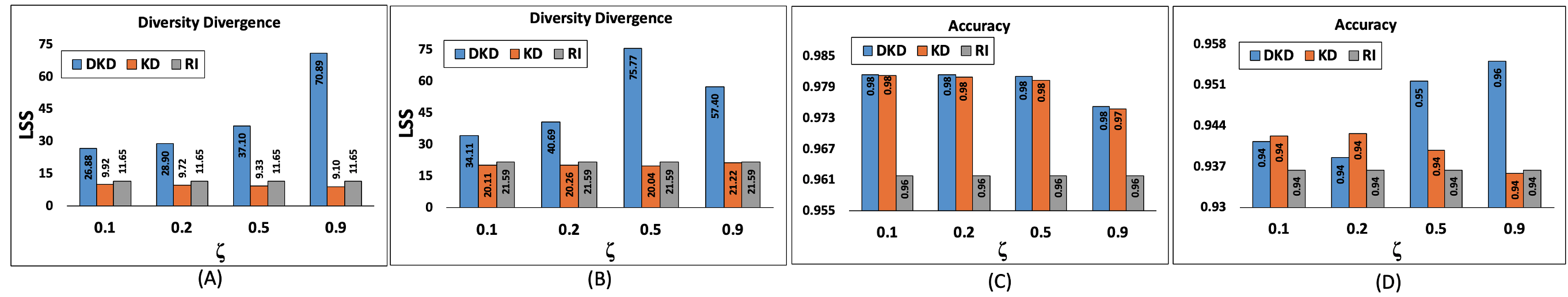}
    
    \caption{(A), (B) show the LSS of an ensemble of three models on the MNIST, and CIFAR10 datasets. (C), (D) show the classification accuracy of the ensemble model on the MNIST, and CIFAR10 datasets. In this figure, three method Random Initialization(RI), Knowlege Distillation (KD), and Diversity Knowledge Distillation(DKD) are shown. }
    \label{ls_gap_real}
\end{figure*}

Fig. \ref{ls_gap_real}-top shows the $LSS$ for an ensemble of three models (for MNIST and CIFAR10 datasets), with the structures described in Table. \ref{architecture}. Fig. \ref{ls_gap_real}-A and C show that for the MNIST dataset, increasing the value of the $\zeta$ causes 1) rapid increase at the $LSS$ 2) slight drop at the classification accuracy. Considering the Eq. \ref{eq:loss_b}, increasing the $\zeta$ means putting less emphasis on the cross-entropy term, which reflects the slight drop of accuracy and increases the $LSS$ of the latent spaces of the ensemble models. A similar pattern happens for the  CIFAR10 dataset, Fig. \ref{ls_gap_real}-B and D, in which $LSS$ increases to its maximum at the $\zeta = 0.5$ while the classification accuracy slightly increases. Between all the different values for acceptable accuracy, the $\zeta$ value that leads to a higher $LSS$ is selected as the parameter. In other words, to have a diverse latent space, the LSS between them should be maximized while the accuracy is kept in the acceptable range. For example, in Fig. \ref{ls_gap_real}-B, when $\zeta$ is 0.5, the LSS is at the maximum level while accuracy is also at an acceptable level. Accordingly, the evaulations has been done when the parameters $\zeta$ is set to 0.5 and 0.9 for the CIFAR10 and MNIST datasets.

One instant observation in Fig. \ref{ls_gap_real} is that the $LSS$ between the latent spaces obtained by DKD approach is noticeably larger than RI, and the $LSS$ between latent spaces obtained by RI is slightly larger than KD. This observation is aligned with our expectations because the DKD is designed to increase the diversity between the latent spaces while the KD in essence, increases the similarity between models and this is because the student model(s) imitates the behavior of the teacher(s).




\begin{table*}
	
	\caption{Black Box adversarial attack on an ensemble of three models. The bold numbers show the most resistant defense mechanism. The perturbation size, $\epsilon$, is set to 0.1 and 0.2 for MNIST, and 0.04, and 0.08 for CIFAR10. The initial constant $C$ is set to 10 and 0.1 for MNIST and CIFAR10 respectively. Iterative attacks are executed for 200 iterations.}
	\label{bb_result}
    \begin{adjustbox}{width=1.8\columnwidth,center}
	\begin{tabular}{|c|c|c|c|c|c|c|c|c|c|c|c|c|c|c|c|}
		\hline 
		\multicolumn{8}{|c|}{MNIST} & \multicolumn{8}{c|}{CIFAR10}\tabularnewline
		\hline 
		\hline 
		\multirow{2}{*}{Param.} & \multicolumn{6}{c|}{3 Ensemble Models} & \multirow{2}{*}{Ref.} & \multirow{2}{*}{Param.} & \multicolumn{6}{c|}{3 Ensemble Models} & \multirow{2}{*}{Ref.}\tabularnewline
		\cline{2-7} \cline{3-7} \cline{4-7} \cline{5-7} \cline{6-7} \cline{7-7} \cline{10-15} \cline{11-15} \cline{12-15} \cline{13-15} \cline{14-15} \cline{15-15} 
		     & KD & KD{*} & DKD & DKD{*} & RI & RI{*} &   &      & KD & KD{*} & DKD & DKD{*} & RI & RI{*} & \tabularnewline 
		\hline 
		\rowcolor{lightgray}
		\multicolumn{8}{|c|}{DeepFool\cite{universal}} & \multicolumn{8}{c|}{DeepFool}\tabularnewline
		\hline 
		1    & 0.9510 & 0.9542	&	0.9721 &	\textbf{0.9726} & 0.962	 & 0.9682  &	0.9754   & 1    & 0.8355	&0.8475	&0.92	&\textbf{0.9463}	&0.8596 & 0.92	&0.6505\tabularnewline 
		\hline 
		200 & 0.864  & 0.8642	&	0.8904 &	\textbf{0.8916} & 0.8433  & 0.8768  &	0.5835  & 200  & 0.7972	&0.8108	&0.8981	&\textbf{0.9269}	&0.8265 & 0.8981	&0.1586 \tabularnewline 
		\hline 
		\rowcolor{lightgray}
		\multicolumn{8}{|c|}{C\&W\cite{cw}} & \multicolumn{8}{c|}{C\&W}\tabularnewline
		\hline 
		1   & 0.9612	& 0.9614 &	0.988  &	\textbf{0.9882} &	0.9721	& 0.9726 &	0.9841 & 1    & 0.8475	&\textbf{0.959}	&0.9534	&0.9575	&0.9326	&0.9571	&0.9516 \tabularnewline 
		\hline 
		200  & 0.7829	& 0.783	 &  0.8587 &	\textbf{0.862}  &	0.8127	& 0.8149 &	0.2 & 200  & 0.8014	&0.9311	&0.9245	&\textbf{0.9321}	&0.9022	&0.9301	&0.1543 \tabularnewline 
		\hline
		\rowcolor{lightgray}
		\multicolumn{8}{|c|}{JSMA\cite{papernot2016limitations}} & \multicolumn{8}{c|}{JSMA}\tabularnewline
		\hline 
		1    & 0.9612	& 0.9614	& 0.9882	& \textbf{0.9884}	& 0.9721	& 0.9726	& 0.9854  & 1   & 0.8475	&0.8599	&0.9326	&\textbf{0.9571}	&0.8348	&0.8703	&0.9516\tabularnewline 
		\hline 
		200  & 0.8147	& 0.8148	& 0.9086	& \textbf{0.9112}	& 0.8403	& 0.8418	& 0.4322 & 200  & 0.7804	&0.7943	&0.8851	&\textbf{0.9145}	&0.7644	&0.8062	&0.1564 \tabularnewline 
		\hline 
		\rowcolor{lightgray}
		\multicolumn{8}{|c|}{FGSM\cite{Goodfellow2014}} & \multicolumn{8}{c|}{FGSM}\tabularnewline
		\hline 
		0.04 & 0.9553	& 0.9555	& 0.9842	& \textbf{0.9846}	& 0.9638	& 0.9644	& 0.952 & 0.04 & 0.8412	&0.8548	&0.899	&\textbf{0.933}	&0.8296	&0.8676	&0.4822 \tabularnewline 
		\hline 
		0.08 & 0.9247	& 0.9249	& 0.9619	& \textbf{0.9629}	& 0.9333	& 0.9341	& 0.864  & 0.08 & 0.7476	&0.7615	&0.7393	&\textbf{0.7763}	&0.7267	&0.7627	&0.2534 \tabularnewline 
		\hline 
		0.1 & 0.8937	& 0.894	    & 0.9374	& \textbf{0.9395}	& 0.9059	& 0.9071	& 0.7847 & 0.1  & 0.6998	&\textbf{0.7131}	&0.6839	&0.7213	&0.6817	&0.7111	&0.0867 \tabularnewline 
		\hline 
		\rowcolor{lightgray}
		\multicolumn{8}{|c|}{No Attack} & \multicolumn{8}{c|}{No Attack}\tabularnewline
		\hline 
		-    & 0.9612	& 0.9614	& 0.9882	& \textbf{0.9884}	& 0.9721	& 0.9726	& 0.9854 & -     & 0.9501	& 0.9512	& 0.9561	& \textbf{0.9580}	& 0.9385	& 0.9523	& 0.9607 \tabularnewline 
		\hline 
	\end{tabular}

	\end{adjustbox}
\end{table*}

\begin{table*}[!htbp]
    \caption{white box attack on the ensemble of three models. Bold numbers at each column show the most resistant method against white-box attacks. The perturbation size, $\epsilon$, is set to 0.1 and 0.2 for MNIST, and 0.04, and 0.08 for CIFAR10. The initial constant $C$ is set to 10 and 0.1 for MNIST and CIFAR10 respectively. Iterative attacks are executed for 200 iterations.}
    \label{white_box}
    \centering
    \begin{adjustbox}{width=1.8\columnwidth,center}
    	\begin{tabular}{|c|c|c|c|c|c|c|c|c|c|c|c|c|}
    		\hline 
    		& \multicolumn{6}{c|}{\textbf{MNIST}} &   \multicolumn{6}{c|}{\textbf{CIFAR10}}\tabularnewline
    		\hline 
    		\hline 
    		\textbf{Defense}                        & \textbf{Clean} & \textbf{FGSM 0.1} & \textbf{FGSM 0.2} & \textbf{JSMA} & \textbf{C\&W} & \textbf{DeepFool} &    \textbf{Clean} & \textbf{FGSM 0.04} & \textbf{FGSM 0.08} & \textbf{JSMA} & \textbf{C\&W} & \textbf{DeepFool}\tabularnewline 
    		\hline 
    		No Defense    &   0.9661    &   0.651   &  0.119 & 0.2421  &  0.4409 &  0.0    &   0.9304  &    0.2033   &   0.1846  &   0.2525  &     0.2548  & 0.1441\tabularnewline         
    		\hline 
    		DKD{*}$_{Prj}$ &  0.9884   &   \textbf{0.9835}  &  \textbf{0.9681}   &  0.8849  &  0.9594  & 0.8991  & \textbf{0.9604}   &  0.7377  &  0.7188  &  0.7788 &  0.8878  & 0.7934 \tabularnewline         
    		\hline 
    		DKD{*}$_{Agg}$  &    0.9884   &  0.9756  &     0.9329  &    \textbf{0.9725}   & \textbf{0.9808}    & \textbf{0.9605}   &   \textbf{0.9604}  & \textbf{0.8923}   &  \textbf{0.8623}  &    \textbf{0.9221}  &   \textbf{0.9538}  & \textbf{0.8751} \tabularnewline         
    		\hline 
    		Yu et al.\cite{yu2018interpreting}   &   0.984   &  0.916  &   0.703 &   0.8014  &  0.791   & 0.6518 &  0.9421 &  0.485 &  0.382   & 0.824    & 0.629  & 0.721 \tabularnewline         
    		\hline 
    		Ross et al.\cite{ross2018improving}  &   0.992   &  0.916  &  0.604  &  0.9191   &  0.753   & 0.7394 & 0.9491  & 0.395  & 0.205    &  0.836   &  0.478  & 0.751 \tabularnewline         
    		\hline 
    		Pang et al.\cite{pang2019improving}  &   \textbf{0.995}    &  0.963   &  0.528 & 0.9465    &   0.781  & 0.7921 &  0.9219 & 0.716   &  0.474     & 0.882    & 0.549  & 0.695 \tabularnewline         
    		\hline 
    		AdvTrain\cite{kurakin2016adversarial}  &   0.991   &  0.73   &  0.527 & 0.645    &   0.392  & 0.627 &  0.905 & 0.446   &  0.314     & 0.781    & 0.501  & 0.73 \tabularnewline         
    		\hline 
    	\end{tabular}
	\end{adjustbox}
\end{table*}

\subsection{Resistance to Black-Box Attacks}
For launching a black-box attack, the adversary uses a reference model and trains it based on the available dataset. Then knowing the transferability of the adversarial example, the adversary extracts the adversarial sample on the reference model and applies it to the models under attack. Assume the adversary used LENET and VGG16  as the reference model for MNIST and CIFAR10, respectively. The underlying models under attack are an ensemble of three models with the structure shown in Table \ref{architecture}. For investigating the performance of the proposed method (DKD) on black-box attacks, we also considered two other methods RI, and KD for training an ensemble of three models.

We used the majority voting between the ensemble models. However, in some cases, each one of the models results in a different prediction regards to other models. We refer to these cases as failed majorities. So for each one of the possible attacks and two datasets (MNIST and CIFAR10) we counted the number of failed majorities. Table \ref{majority_fail}, shows the difference between the regular and boosted version of each benchmark with regard to the number of the failed majority voting. From this table, we observe that 1) the number of majority voting failures at the DKD is always higher than the other two regular methods. This confirms that our objective function could successfully train diverse models because the majority voting fails whenever the models couldn't agree on a label. So at the presence of the adversarial example, the disagreement between the models trained with DKD is higher than the others 2) The number of majority failures drops by going from a regular to a boosted model. The effect of this drop is shown by Accuracy Improvement percentage, A.I. 

Table \ref{bb_result} shows the results of applying some of the state-of-art attacks on the DKD, KD, and RI on the  MNIST, and CIFAR10 datasets. Investigating these two Tables two trends reveals 1) Boosted version has better performance than the regular version 2) Almost at all attack scenarios, $DKD^{*}$ outperform the other defense scenarios.

\begin{figure}
    \centering
    \includegraphics[width=0.9\columnwidth]{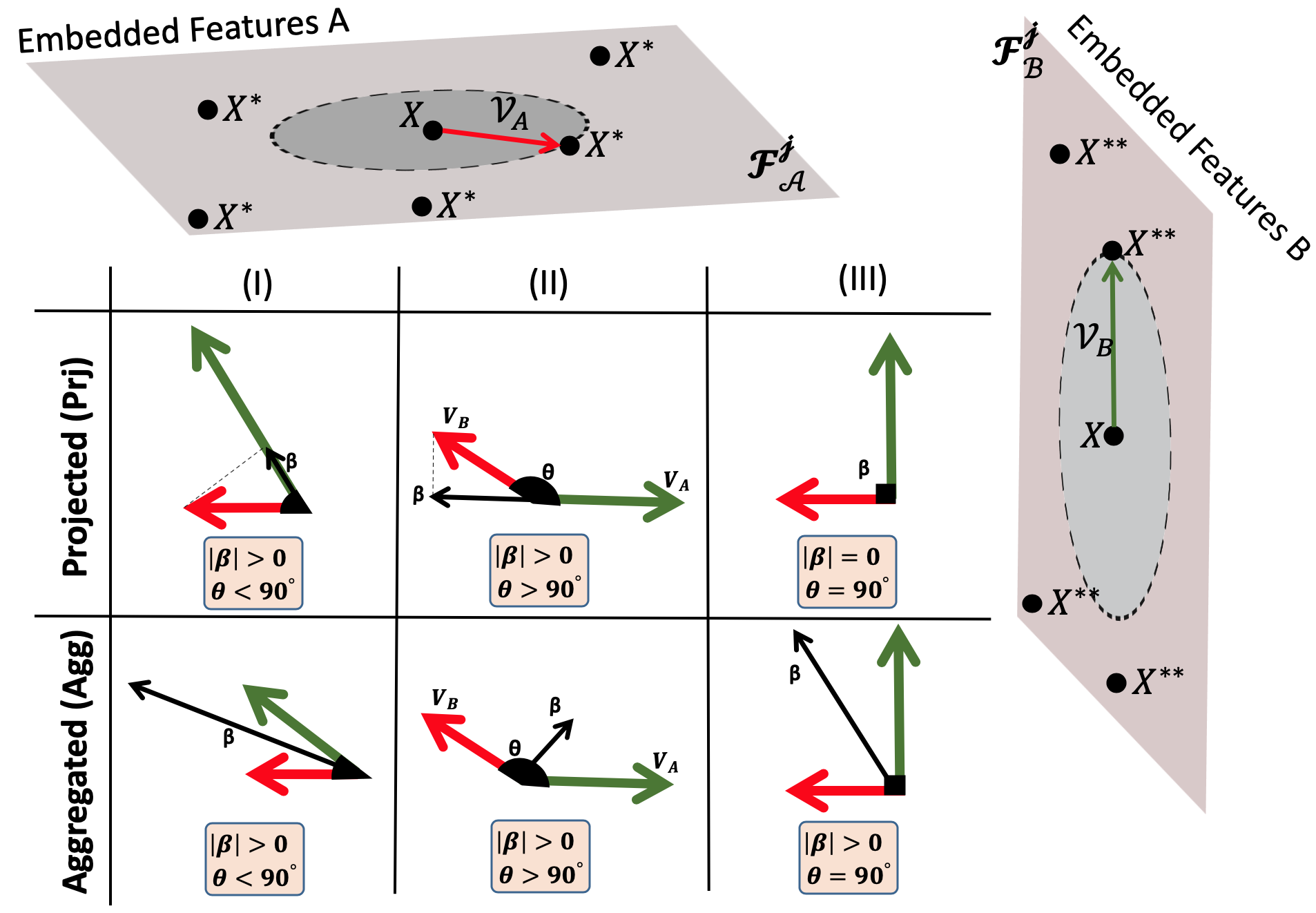}
    \caption{White Box attack scenarios for an ensemble of two models A and B in which $\mathcal{F}_\mathcal{A}^{j}$ and $\mathcal{F}_\mathcal{B}^{j}$ are corresponding embedded features for layer j$^{th}$ of each model respectively. $\mathcal{V}_A$ and $\mathcal{V}_B$ are the adversarial perturbations on input sample $x$, i.e., $X^* = \mathcal{F}_{\mathcal{A}}^{{j}}(x) + \mathcal{V}_A$ and $X^{**} = \mathcal{F}_{\mathcal{B}}^{{j}}(x) + \mathcal{V}_B$. The angle between two vectors $\mathcal{V}_B$ and $\mathcal{V}_A$ is annotated with $\theta$, while $\beta$ is the magnitude of projected or aggregated perturbation for Projected or aggregated attack, respectively. }
    \label{space}
\end{figure}

\subsection{Resistance to White-Box Attacks}

We evaluated 2 white-box attacks: 1) Projected attack, in which the adversary obtains adversarial perturbation on one model and applies it to the target model(s). 2) The Aggregated attack, in which the adversary obtains the adversarial perturbation on each model independently, then aggregates them to form one perturbation to apply to all models simultaneously. These two scenarios have been explained through a toy example on two models $A$ and $B$ in Fig. \ref{space}. For the projected attack, the adversary finds the direction $\tau_B$ as an adversarial perturbation for the input $x$ of the model B, i.e., $\mathcal{F}_{\mathcal{B}}^{j}(x +\tau_B) = X+\mathcal{V}_B = X^{**}$. In the second step, the adversary applies the obtained perturbation $\tau_B$ on the sample of the model A, with the hope that it may transfer to model A. Noted, projected attack is similar to black box scenario with only difference that attacker knows about the structures of auxiliary models and in this paper the auxiliary models have a same structure. Hence, based on the angle between the perturbations in the latent spaces of models A and B which are $\mathcal{V}_A$ and $\mathcal{V}_B$ three scenarios are possible, see Fig. \ref{space}. When both perturbations $\mathcal{V}_A$ and $\mathcal{V}_B$ are orthogonal (case III), a successful adversarial perturbation of one model does not transfer to the other model, and vice versa. In two other cases (I, II), the smaller the perturbations' angle, the more likely the adversarial perturbation transfer across models. Note that the projection of the perturbation $\tau_B$ on the embedded feature plane $A$, $\mathcal{F}_{\mathcal{A}}^j(\tau_B)$,  shows the direction of the adversarial perturbation for the model A. If this projection is large enough to move the data point $X$ out of its class boundary then input $X$ misclassifies as $X^{*}$.


For the aggregated attack, the adversary obtains the adversarial perturbation of input sample $x$ on the model A, $\tau_A$, and separately obtains the adversarial perturbation  B, $\tau_B$. Then, the adversary calculates the aggregated perturbation by adding up these two perturbations i.e., $\tau_A + \tau_B$. In this scenario, let's assume $\mathcal{V}_A$, and $\mathcal{V}_B$ are the corresponding mapping of $\tau_A$ and $\tau_B$ on embedded feature planes A, and B, respectively. Based on the value of $\theta$ between $\mathcal{V}_A$, and $\mathcal{V}_B$, three different outcomes can be assumed, 1) Fig.\ref{space} Aggregated-II, $\theta > 90$, in which the aggregated perturbation i.e., $\beta$ is smaller than either of perturbations $\mathcal{V}_A$, and $\mathcal{V}_B$. In this case, $\beta$ as an adversarial perturbation cannot be a successful attack on either of the models. 2) Fig.\ref{space} Aggregated-I, $\theta < 90$, in which  $\beta$ is greater than either of perturbations $\mathcal{V}_A$, and $\mathcal{V}_B$, which means $\beta$ as an adversarial perturbation leads to a misclassification in both models A, and B. However the resulted adversarial perturbation in this scenario is large and most likely perceptible to human. 3) Fig.\ref{space} Aggregated-III, $\theta = 90$, in which $\beta$ is equal to either of perturbations $\mathcal{V}_A$, and $\mathcal{V}_B$, which means $\beta$ as an adversarial perturbation lead to a misclassification on both models A, and B  and most likely $\beta$ is imperceptible to the human eyes.

Table \ref{white_box} captures the result of various adversarial attacks on our proposed solution. DKD$^{*}_{Prj}$ shows the evaluation of DKD{*} when the adversary uses the projected attack and DKD$^{*}_{Agg}$ shows the aggregated attack. As indicated in this table, our proposed solutions outperform prior art defense, illustrating the effectiveness of learning diverse features using our proposed solution.

\section{Acknowledgement}
This work was supported by Centauri Corp. and the National Science Foundation (NSF) through Computer Systems Research (CSR) program under NSF award number 1718538. 

\section{Conclusion}
To build robust models that could resist adversarial attacks, we proposed a solution for building an ensemble learning solution in which member models are forced to extract different features and learn radically different latent spaces.  We also introduced Latent Space Separation (which is defined as the distance between the latent space representations of models in the ensemble) as a metric for measuring the ensemble's robustness to adversarial examples. The evaluation of our proposed solutions against the white and black box attacks indicates that our proposed ensemble model is resistant to adversarial solutions and outperforms prior art solutions. 

\renewcommand{\IEEEbibitemsep}{0pt plus 0.5pt}
\makeatletter
\IEEEtriggercmd{\reset@font\normalfont\fontsize{7pt}{6.5pt}\selectfont}
\makeatother
\IEEEtriggeratref{1}

\bibliographystyle{IEEEtran}
\bibliography{main}

\end{document}